\documentclass[]{aa}
\usepackage{graphicx}
\usepackage{txfonts}
\usepackage{natbib}
\bibpunct{(}{)}{;}{a}{}{,} 

\title{Test particle acceleration in \\ a numerical MHD experiment of an anemone jet}

\author{Karl Joakim Rosdahl$^{1,2}$ \and Klaus Galsgaard$^2$}

\institute{CRAL - Observatoire de Lyon, 9, avenue Charles Andre, 69561 Saint Genis Laval cedex, France 
\and 
Niels Bohr Institute, Blegdamsvej 17, Dk-2100 Copenhagen {\O}, Denmark}

\date{\today}

\abstract{}
{To use a 3D numerical MHD experiment representing magnetic flux emerging into an open field region 
as a background field for tracing charged particles. The interaction between the two flux 
systems generates a localised current sheet where MHD reconnection takes place. We 
investigate how efficiently the reconnection region accelerates charged particles and what
kind of energy distribution they acquire.}
{The particle tracing is done numerically using the Guiding Center Approximation 
on individual data sets from the numerical MHD experiment.}
{We derive particle and implied photon distribution functions having power law forms, 
and look at the
impact patterns of particles hitting the photosphere. We find that particles reach
energies far in excess of those seen in observations of solar flares. However the 
structure of the impact region in the photosphere gives a good representation of the 
topological structure of the magnetic field.}
{}

\keywords{Acceleration of particles -- Magnetic fields -- Magnetohydrodynamics (MHD) -- Sun: corona}

\authorrunning{Rosdahl \& Galsgaard}
\titlerunning{Particle acceleration in a numerical MHD experiment}

\newcommand{\Eq}[1]{Eq.~(\ref{#1})}
\newcommand{\Fig}[1]{Fig.~\ref{#1}}
\newcommand{\Sec}[1]{Section~\ref{#1}}
\newcommand{\EQ}{\begin{equation}}  
\newcommand{\EN}{\end{equation}}
\newcommand{\EQA}{\begin{eqnarray}}
\newcommand{\ENA}{\end{eqnarray}}
\def\BB {{\bf  {B}}}
\def\EE {{\bf  {E}}}
\def\vv {{\bf  {v}}}
\def\pp {{\rm  {p}}}
\def\RR {{\rm  {R}}}

\begin{document}

\maketitle

\section{Introduction}
To investigate the dynamical evolution of the large scale phenomena seen
in the solar atmosphere, one traditionally adopts the magneto-hydrodynamic (MHD) 
representation. The MHD approach is a macroscopic
approximation describing the time dependent evolution of a plasma. The assumption
when adopting MHD is that the time dependent plasma evolution is well described by
macroscopic parameters, such as temperature, density,
bulk velocity and magnetic field. This requirement is fulfilled when the plasma is 
in a local thermodynamic equilibrium and the characteristic MHD length scale is much
longer than the mean free path of the plasma particles. 
For most of the time the plasma in the lower solar 
atmosphere obeys this requirement, but there are special situations 
for which MHD is not a good representation. One such situation 
is magnetic reconnection. Despite this fact, 
magnetic reconnection has for many decades been investigated using the MHD 
approximation \citep[See][for a review.]{2000mare.book.....P}. 
In MHD the non-ideal evolution imposed by viscosity and resistivity in the plasma is 
described through a parameterization that approximates the real particle interaction on
length scales well below the typical resolution of MHD. The choice of the form and 
value of the magnetic resistivity, $\eta$, naturally has implications on how the 
reconnection process progresses. 

From the vast literature on 2D MHD reconnection, a common feature is the presence
of a localised diffusion region, from which standing slow mode shocks extend 
along the separator lines -- the Petschek like picture \citep{1964NASSP..50..425P}.
When redoing this type of experiments with particle codes, one finds a
different picture. The existence of the MHD slow shocks vanishes as the
evolution of protons and electrons decouples on sufficiently short length scales
removing the plasma's ability to make a single coherent shock structure 
\citep[See the article by Drake and Shay]{2007rmfm.book.....B}.
Particle simulations allow for distribution functions to reach significant
deviations from Maxwell distributions, typically providing a power law tail of high energy
particles accelerated away from the diffusion region. One could therefore fear
that the global evolution of the field will be qualitatively different depending on
which method one chooses to use for the investigation.
Recent comparisons between different approximations, going from MHD to particle in cell (PIC), 
in a controlled reconnection experiment have 
shown that despite variations in the reconnection speed between the different approaches,
the same amount of flux reconnects and the same general structure
of the final magnetic field results \citep{2005GeoRL..3206105B}. Numerical MHD experiments
therefore do capture the large scale dynamics of the evolution, and will at worst provide
a somewhat wrong time scale and energetics for the dynamical evolution.

In the Solar atmosphere magnetic reconnection is deemed responsible for various explosive
events, and the conversion of magnetic energy into other forms is therefore
an important process one needs to understand to be able to explain observations.
In the MHD approximation magnetic energy is converted to bulk motion and Joule heating
of the plasma. This is a limitation, that already became clear from observations of 
X-rays and $\gamma$-rays in the late 1970's. Already then it was found that a significant 
amount of the energy release associated with large flares is converted into energetic 
particles (10-50\%) \citep{1976SoPh...50..153L,2003ApJ...595L..69L}.
The effect of the accelerated particles is seen in different characteristic features.
In the general picture one assumes that reconnection takes place in the corona, and
that the strong electric field parallel to the magnetic field accelerates electrons and
protons to velocities where their cross section becomes so small that they
interact very little with the coronal plasma. Only when they reach the dense plasma
close to the photosphere are they stopped very efficiently while, causing them to emit
bremsstrahlung radiation.
In loop flares, this gives rise to $H_{\alpha}$ ribbons in the photosphere that move apart
with time. The combined information of the motion and the magnetic field distribution in
the photosphere provides indirect information about the reconnection speed in the 
acceleration region
\citep{1981ApJ...246L.155H,2002ApJ...566..528I,2007ApJ...664L.127J,2008ApJ...672L..69L,2009ApJ...696L..27L}.
In the reconnection model, one also expects the particles to be accelerated along the open 
field lines reaching into the upper corona. Hard X-ray observations have been made at the 
loop top of flares \citep{1994Natur.371..495M}, but it is not until recently that RHESSI 
has been able to detect a signature of independent particle distributions on either side 
of an assumed current sheet 
\citep{2008ApJ...678L..63K}. This investigation strongly supports the picture where the 
acceleration region is located in the corona, and that particles are accelerated
in both directions away from the local reconnection region. 
Their analysis also indicate that the
very energetic particles can be maintained in the local vicinity of the diffusion region
long enough to be able to interact with the local plasma before continuing away from
this region. Understanding the particle acceleration and interaction process is 
therefore important for understanding the distribution of the released magnetic energy in
the flare process.

The lack of information regarding particle acceleration in numerical MHD experiments has led us
to start investigating its effect. We adopt a simplified approach, where test particles 
are traced
in snapshots from a numerical MHD experiment. The aim is to investigate the efficiency
of the acceleration through forming distribution functions, looking at the patterns
they form as they exit the allowed tracing domain and especially calculate radiation
spectra from the bremsstrahlung that arises when the fast electrons are stopped in 
the top layer of the dense photosphere. 

The layout of the paper is as follows. 
In the following section we review the relevant information about the numerical MHD experiment used
as background field for test particle tracing. \Sec{Code.sec} describes the numerical 
approach used for tracing the particles, while \Sec{Setup.sec} gives the setup of the
tracing experiment. In \Sec{Results.sec}, the results of the
data analysis are presented. \Sec{Discussion.sec} discusses the drawbacks 
with the approach and what we need to do to improve the results, while \Sec{Conclusions.sec} 
draws up the conclusions of the paper and suggests further work to be done.

\section{Numerical MHD experiment}
\label{Experiment.sec}

\begin{figure}
  {\hfill \includegraphics[width=0.35\textwidth]{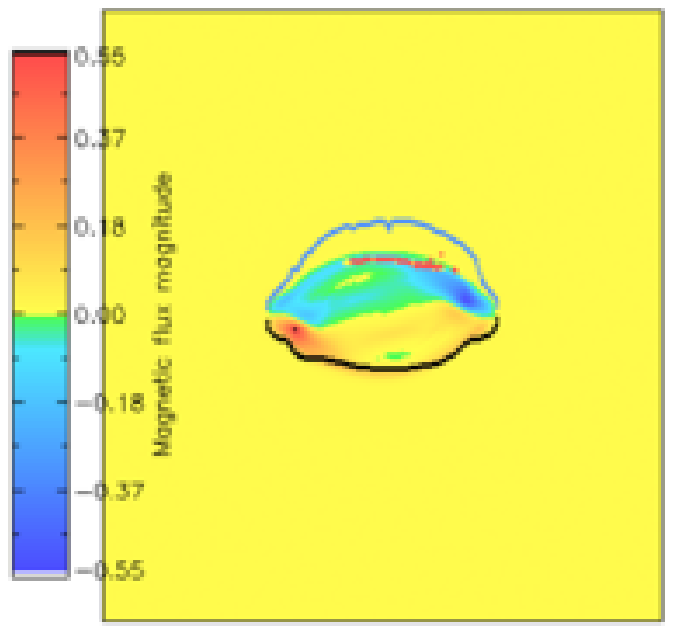} \hfill}

  {\hfill \includegraphics[width=0.35\textwidth]{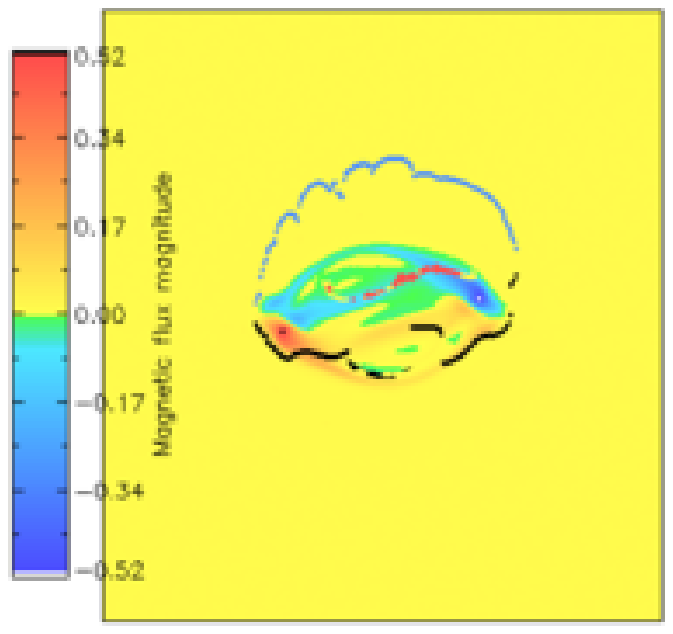} \hfill}

  {\hfill \includegraphics[width=0.35\textwidth]{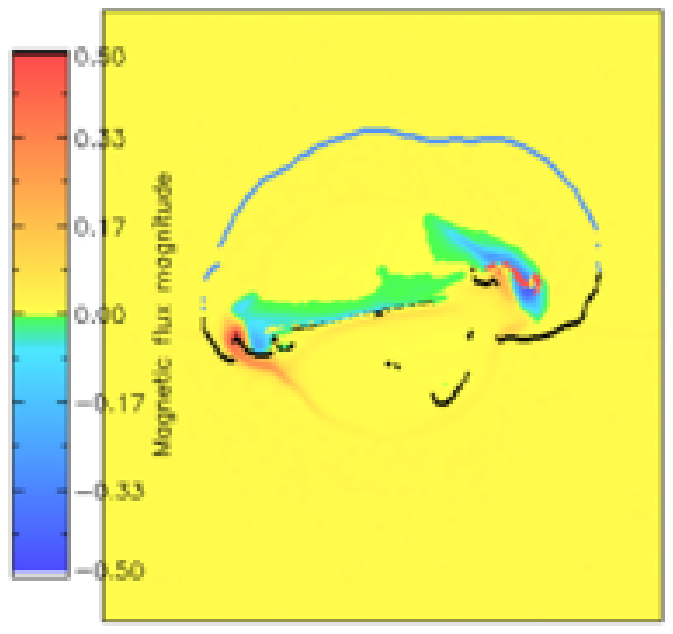} \hfill}

  \caption[]{\label{connectivity.fig}The connectivity pattern of the magnetic field 
    in the emergence process at three different times. The colour indicates the field 
    strength and sign of the normal component of the magnetic field with blue representing
    negative flux. Each frame is scaled to its own dynamical range. The lines indicate the 
    intersection of the separatrix surfaces with the photosphere.}
\end{figure}

The dynamical evolution of an emerging flux tube entering into an open coronal field 
was investigated using a numerical 3D MHD approach by \citet{2008ApJ...673L.211M}.
The numerical experiment consists of a stratified atmosphere covering the region from
4 Mm below the photosphere to 29 Mm above it. Imbedded in the convection
zone is a twisted flux tube that over a fraction of its length is made buoyant.
Additionally a slanted magnetic field permeates the numerical domain, representing
an open magnetic field configuration. The part of the flux tube that rises from the
convection zone pushes away the slanted magnetic field to make space for itself. While 
the tube is still in the convection zone, the motions are slow and not very compressible, 
and very little interaction between the two flux systems takes place. 
As the rising tube reaches the photosphere the magnetic field strength starts to
build up. Shortly after, the vertical magnetic pressure gradient becomes sufficient to
initiate the emergence of the flux tube into the upper atmosphere. The emergence speed
in this phase is much higher than the slow rise speed observed in the convection zones. 
The initial plasma $\beta$ decreases with height through the transition region and into the
corona making it easier for the two magnetic flux systems to come into close contact.
The change in $\beta$ is important for the dynamical evolution of the boundary layer
between the two flux systems, especially where they experience large 
differences in field line orientation. Such regions are prone to generating strong 
current concentrations as they are compressed by the emergence process. As part of 
the flux tube emerges into the corona a current 
dome separating the two flux systems is formed. The continued emergence concentrates the 
current in the dome into a long thin sheet connecting along the length of the emerged 
flux, laying skew 
to one side of the summit line of the emerged flux. Eventually the current density becomes 
large enough for non-ideal MHD processes to start changing the field line connectivity. The 
reconnection between the magnetic field from the rising loop and the ambient open field
creates a new low lying loop system next to the emerging one and new field lines that
connect the deeper parts of the twisted loop with the open flux. The effect of reconnection
is therefore to eat away the emerging flux and build up a secondary loop system. One way 
to see this is to investigate how the separatrix surface between the two flux systems changes 
with time. This is investigated by Galsgaard \& Moreno-Insertis (2010, in prep.) and shows how 
the interaction between the magnetic field and the photosphere changes structurally with time,
clearly indicating the continued diminishing 
of the area occupied by the emerging loop system and the concurrent growth of the new loop
system. \Fig{connectivity.fig} shows the connectivity domains
at three different times during the emergence process. The background shading represents the
photospheric magnetogram indicating the sign and magnitude of the magnetic fields normal
component. This shows a bipolar structure imbedded in a monopolar background field. 
From the magnetogram alone one could be led to conclude that a new bipolar field has emerged 
into the photosphere with no complicated connectivity pattern. This is not the case: The lines 
on the map show the intersection
of the separatrix surfaces with the photosphere. The main features are the outer line defining
the perimeter inclosing flux that connects both ends to the photosphere.
A line dividing the perimeter represents the division between the two loop 
systems, where the top one is the secondary loop created by the magnetic reconnection process.
The jagged form of the outer line is created by convective motions that overshoot above
the plane in which this surface is shown.

As a consequence of the changes in magnetic connectivity, two reconnection jets are 
accelerated out of the current sheet edges. The most prominent 
of the jets escapes upwards, giving rise to high velocity plasma motions that follow the open 
field lines away from the solar surface. 

The evolution of the numerical MHD experiment resembles the anemone jets which Hinode observations have
found to be very common in open flux regions \citep{2007Sci...318.1591S}.

The MHD approximation assumes the plasma to be in a statistical equilibrium, defined by
frequent collisions between particles, where physical variables like temperature and
bulk velocity are represent for the average particle in a small volume.
Therefore, numerical MHD experiments are not able to provide indications as to how 
individual particles may react to strong electric fields parallel to the magnetic field.
In a real plasma, this electric field component is responsible for accelerating 
a small fraction of particles up to relativistic velocities. As only a small fraction of
the particles are accelerated, they give rise to a high energy power law tail attached to
the Maxwellian velocity
distribution. Power law distributions created in eruptive events like solar flares 
are well documented by RHESSI observations 
\citep{ 2008ApJ...676..704L, 2008AIPC.1039...52L, 2009ApJ...696L..27L}. 
In fact, integrating the energy content of this tail indicates that a large fraction, 
up to 50\%
\citep{2002ESASP.506.1035L,2003ApJ...595L..69L}, 
of the energy released in the flare process may go into accelerating particles.
If we therefore want to understand in detail the effect of the flare processes
it is important to take this additional physics into account. This
is not an easy task, as it requires models to resolve particle interactions, while still 
maintaining the large scale structures responsible for the dynamical driving of the magnetic 
system. Resolving both the particle scale and the large structural scales requires
unrealistically large numerical resolution. A simpler approach is to assume:
\begin{itemize}
  \item{The integration time for the particle experiment is much shorter than the time 
        between MHD snapshots.}
  \item{The MHD snapshot is static in time.}
  \item{There is no back reaction on the MHD field from the particle motions.}
\end{itemize}
The first assumption is fulfilled in the following analysis, while the two later assumptions are 
unrealistic. However, using these makes it possible to obtain results regarding (i) the 
particle distributions after acceleration, (ii) the locations where 
the acceleration takes place and (iii) where the accelerated particles end up as they 
are rapidly accelerated away from their initial positions. The details of the applied 
approximation for this approach are discussed in the following section.

\section{{\bf Test} particle tracing approach}
\label{Code.sec}

The relativistic equation of motion for a charged particle in the presence of a Lorentz force
is described by the Lorentz equation (LE):
\EQ
{{d \pp} \over {dt}} = q\left( \EE + \vv \times \BB \right),
\label{Lorentz.eq}
\EN
where $\pp = m \gamma \vv$ is the relativistic momentum, $q$ the particle charge, $\EE$ and
$\BB$ the electric and magnetic field at the particle position, $\vv$ is the
particle velocity, $m$ is the rest mass of the particle and $\gamma$ 
is the Lorentz factor.
This equation describes the orbiting motion of a charged particle around any magnetic field. 
Before going further into the discussion of the applications of this equation it is worth 
looking into the characteristic radius of the orbiting motion of a charged particle in 
a magnetic field. 
The Larmor radius of this circular orbit is 
\EQ
R_c = {{\gamma m v} \over  {Bq}},
\EN
where $v$ is the speed. For the emergence experiment the characteristic coronal values are
$B=10^{-2}T$ and the coronal temperature 1.2 MK. Assuming a Maxwell-Bolzman distribution
for the particles gives typical electron velocity of roughly $10^7 m/s$ and a $\gamma=1$. 
Using these values in the equation above, we obtain a 
Larmor radius of order $10^{-3}m$. This is a typical value, but more 
extreme circumstances, may change it by up to 2 orders of magnitude. The relevant 
quantity to compare to this value is the resolution of the emergence experiment. In this 
case the box size is $34 Mm$ and the grid resolution on the order of 320 grid points. 
This gives a typical grid resolution in the numerical domain of order $0.1Mm$. Comparing 
numbers it is seen that the orbiting motions typically take place on $10^{-7}$ of a grid 
cell. Obtaining the magnetic field with this spatial resolution, from a single 
precision dataset, implies that the magnetic field seen by a particle in a given orbit,
will be constant across the orbit, and its orbit will only change due to changes in the 
magnetic field direction taking place over large distances measured in Larmor radii. Using 
\Eq{Lorentz.eq} to track particle motions in the numerical MHD experiment will therefore require a 
huge amount of computing time, while providing information that can be obtained using a 
simpler approximation. Instead of using the Lorentz equation to determine the particle paths, 
we use the Guiding Center Approximation (GCA) \citep{1964AmJPh..32..807N}. The GCA
integrates the $\BB$-parallel motion and the non-parallel of charged particles,
while ignoring the orbital motion, but keeping track of the magnetic moment. Ignoring the 
orbital motion implies that the time step for this approach can be up to several orders of 
magnitude larger than for
the Lorentz equations, which can save a huge amount of computing time and provide the same 
particle trajectories. The relativistic GCA equations of motion are:
\EQA
{d u_{||} \over {dt}} & = &  {q \over m} E_{||} - {{\mu} \over {\gamma m B}} \left( \BB \cdot \nabla\right) B, \\
{{d \RR} \over dt} & = &  { u_{||} \over {\gamma B}} \BB + {{\EE \times \BB} \over B^2} + 
                   {\mu \over {\gamma q B^2}} \BB \times \nabla B \\ \nonumber
                   & & + {{m u_{||}^2 } \over {\gamma q B^4}} \BB \times ( \BB \cdot \nabla) \BB, \\ 
\mu & = & {{mu^2_{\perp}} \over {2B}} = constant
\ENA

where the $_{||}$ refers to the component along the magnetic field, $R$ is the center 
coordinate of the particle orbit and $_\perp$ refers to the component perpendicular to 
the magnetic field vector -- here only the magnitude is interesting and not it's angle.
These equations represent the momentum equation, the motion of the guiding center and 
the adiabatic invariance of the magnetic moment. 

We solve these GCA equations numerically using a 5th order Cash-Karp Runge-Kutta 
\citep{1992nrfa.book.....P} method 
with adaptive time stepping. As the electric and magnetic fields are provided by the 3D numerical MHD 
experiment, these values are only known at the grid locations provided by the MHD experiment. 
Values at the actual particle positions are obtained using a Bi-cubic interpolation scheme.

The data from the MHD simulations is saved in dimensionless units, and are scaled to SI units
before being used in the tracing code. The electric field is not saved in the MHD code, 
and one can use different methods to derive the resistive contribution from the saved data 
Numerical codes typically uses resistivity models at constant $\eta$, but with various 
forms of current dependent $\eta$ threshold, higher than 2nd order current dependence -- typically
4th order -- to localise the effective $\eta$ towards the numerical resolution limit combined with 
a "shock" capturing approach to localise diffusion in current sheet regions as long as new 
magnetic flux is advected into the sheets.
In this case we have used the algorithm that was used in the numerical MHD experiment. The MHD code 
uses a high order finite difference approach on staggered grids to solve the MHD equations. 
The high order approach (6th order derivatives
and 5th order interpolations) requires the special approach with 4th order dependence combined with
the "shock" capturing mechanism to avoid/limit numerical ringing effects
in the physical variables in the vicinity of regions where the magnetic field changes 
significantly over a few grid points. A description of the 
algorithm is given in \citet{Nordlund+Galsgaard95mhd}.

\section{Particle setup}
\label{Setup.sec}

Test particle tracing has previously been done in a number of different environments. Mostly
these have consisted of situations representing MHD turbulence 
\citep{2003ApJ...597L..81D,2004ApJ...617..667D,2006JGRA..11112110D,2006A&A...449..749T}
finding the accelerated particles to reach high velocities, having power law distributions with 
steep power indexes \citep{2006A&A...449..749T}. The typical result is that the fastest 
particles reach energies much higher than found in solar observations.

\citet{1988JGR....9314383A} showed, using a 2D reconnection experiments, that having
a single large scale diffusion region with internal structure gives rise to a stochastic acceleration
process where initial close neighboring particles can experience very different accelerations
histories as their paths diverge exponentially with time. This is clearly different from earlier
experiments that consisted of laminar diffusion regions. The same process is seen in
the previous 3D papers 
\citep{2006A&A...449..749T,2003ApJ...597L..81D,2004ApJ...617..667D,2006JGRA..11112110D}, 
where, as a general concept, the whole numerical domain is in a turbulent state with acceleration 
sites scattered randomly throughout the numerical domain.
In the present case we have a situation where the acceleration region is localised in the 
numerical domain, and the global topology of the magnetic field is not in a turbulent state
as in the earlier mentioned 3D experiments. This does not imply that the diffusion region 
has a simple laminar structure, as it contains significant variations in both space and
time and we therefore expect the acceleration process to be stochastic in nature.
Therefore, it is interesting 
to investigate how particles are accelerated in this environment, which type of particle 
distribution arises, where they preferentially "exit" the free flight domain 
and which type of observational spectra are obtained from the bremsstrahlung generated 
as the particles impact with the photosphere.

\begin{figure}
 {\hfill \includegraphics[width=.45\textwidth] {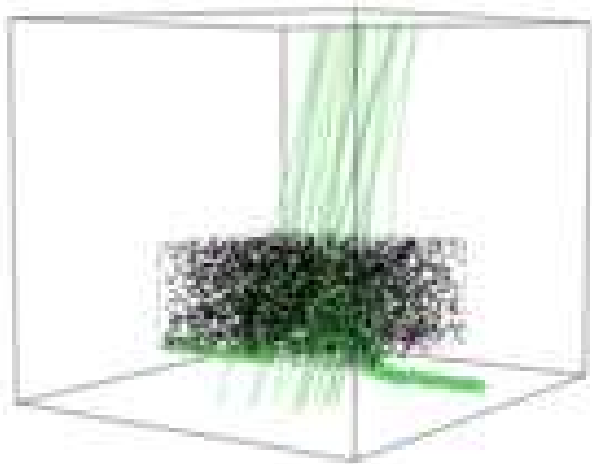} \hfill}
 \caption[]{\label{init_box.fig} The image represents the magnetic field line structure at 
    one instance in the MHD simulations, showing the general structure of the magnetic 
    field by tracing a number of representative magnetic field lines. The dots represent 
    the  starting positions of a smaller sample of the particles in the tracing experiments. }
\end{figure}

To initiate the investigation, a sub volume of the MHD box is chosen to contain the 
traced particles. The lower boundary of this domain is the photosphere in the MHD model, below 
which the plasma density becomes so high that collisions will thermalize any accelerated 
particles almost instantaneously. The top boundary is chosen such that for all the snapshots 
in the numerical experiment it is above the current sheet region that is responsible for
accelerating the particles. The extent along and across the loop is chosen to 
cover the maximum extent of the current sheet. A representation of
the particle box seen in relation to the magnetic field configuration in the emergence phase 
is shown in \Fig{init_box.fig}.

We perform tracing analysis for all snapshots provided by the numerical MHD emergence experiment. 
For each of the 160 snapshots we initiate 90,000 particles in the chosen sub domain. 
Inside this domain, they are given random positions and pitch angles. Finally they are
assigned a velocity according to a Maxwellian distribution with a temperature of 1.2 MK.
The particles are then traced for 0.5 seconds or until they either impact with the 
photosphere or exit the full MHD box above the photosphere. 

Tracing calculations are conducted for both protons and electrons. The results show that the
parallel resistive electric field only provides for a small acceleration of the protons, 
while the much lighter electrons experience acceleration, even on multiple sites, that 
changes the distribution function significantly within the alloted tracing time.  The 
following discussion will therefore only concern the results from the electron experiments.

\section{Results}
\label{Results.sec}

In this section we discuss the results obtained from {\bf test} particle tracing in the static 
MHD snapshots. There are three objectives to this section. The first is to follow the 
particle trajectories to see where they end up after the 0.5 seconds. As a fraction of 
these reach the photosphere, spatial distribution pattern indicates what we may expect to 
see in real observations, while the ones exiting the top boundary are expected to enter into 
the solar wind particle population. The second is to look at the 
velocity distribution functions obtained as a function of MHD snapshot to see how large a 
fraction of the particles are being accelerated and which energies they reach. Finally, we use 
the distribution function for the electrons that impact with the photosphere to derive spectral
information that can be compared directly with observations.

\subsection{Impact patterns}

Following 90,000 particle paths through the MHD domain is 
by no means feasible, but provides detailed insight to the individual reasons for particle 
acceleration. We will look at a few examples of this below. Apart from this, there 
are two other approaches we are going to investigate in this section. One is to 
follow a sample of particles as they are moving around the domain. The most interesting are 
the ones that impact either the photospheric or the top boundary. These give us 
an idea about the locations where we would expect to see secondary effects of the reconnection 
process, and allows us to obtain distribution functions and spectra of the impacting particles.

\subsubsection{Individual acceleration patterns}

\begin{figure}
  {\hfill \includegraphics[width=.45\textwidth] {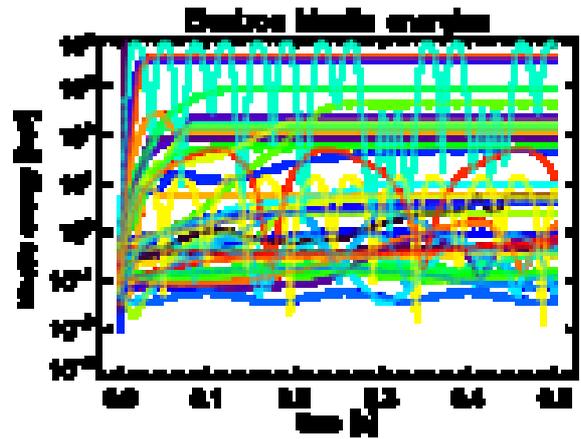} \hfill}
  \caption[]{\label{particle_path.fig} The graph represents the kinetic energy of a 
  small number of significantly accelerated particles as a function of time. 
  The different colours represent
  different particles. The peak energy in the diagram, $10^4$ keV does not encompass the
  most energetic particles in the experiment. It is noticed that all example particles are
  accelerated to energies many orders of magnitude above their thermal energies. About
  half of the sample particles stays at thermal energies.
 }
\end{figure}

In previous papers describing particle acceleration, it has been seen that the 
particles undergo repeated acceleration and de-acceleration as they pass through the 
turbulent magnetic field domain 
\citep{2004ApJ...617..667D,2006JGRA..11112110D,2006A&A...449..749T}.
In this experiment we assume
that a single coherent current sheet is more likely to only provide one single 
acceleration of the particles, at least for the ones that very rapidly reach the
domain boundaries. \Fig{particle_path.fig} shows the time evolution of a number of 
electrons. 
This shows a time history that is similar to those seen in the previous papers,
starting with a rapid acceleration followed by a period of cyclic behavior in 
kinetic energy. We can conclude that there are a number of particles that are
trapped on trajectories where they repeatedly are moving forward and backward along the same
magnetic field line. A small number of particles are seen to reach a peak energy level, 
which is subsequently represented by a line with constant energy. These are particles that
have reached one of the imposed boundaries, from where further tracing is not possible.
As can be seen, this type of particles reach the boundary rather fast, having
experienced a rapid acceleration over a small fraction of a second.

From inspection of the animations of the particle motions (movies are available with the 
Web version of the paper)
it is easy to see both the fast moving particles that exit the domain
and the ones that seem to oscillate around an equilibrium position.


\subsubsection{The global view}

The behavior of the test particles depends strongly on the time of the snapshot from the numerical  MHD 
experiment. Initially the particles are placed in a domain where no interaction between the 
two flux systems takes place. No parallel electric fields exist and the particles are just 
thermally moving along the magnetic field lines without significant changes in their energy. 

\begin{figure}
  Â
  {\hfill \includegraphics[width=0.38\textwidth]{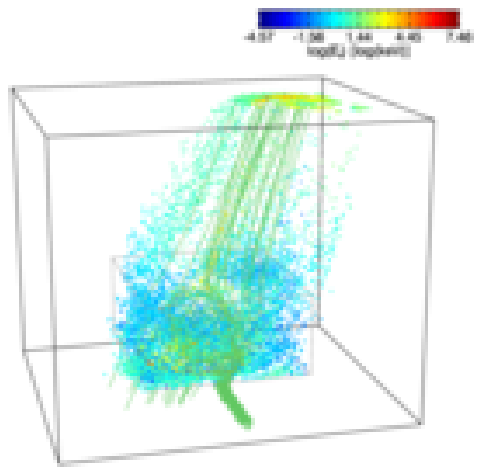} \hfill}

  {\hfill \includegraphics[width=0.38\textwidth]{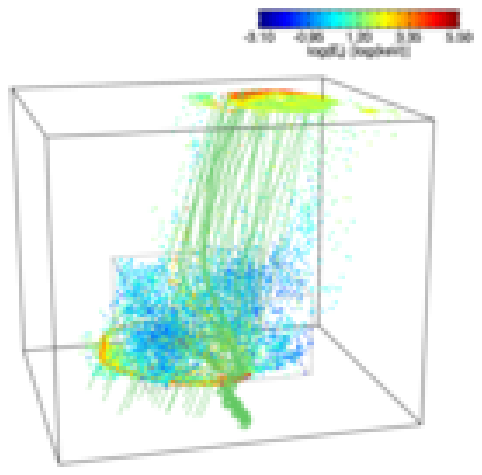} \hfill}

  \caption[]{\label{Trace.fig} The frames show the large MHD domain, with the imposed 
  particles at their final positions for two different times in the numerical MHD experiment. The 
  field lines outline the general structure of the magnetic field. The particles are 
  colored relative to their final kinetic energy.}
\end{figure}

As the flux tube starts to rise into the region where the particles are placed, effects on the 
particle behavior become visible. First the particles stay almost confined to the box in 
which they are initiated, but later they start moving around within this region gaining an 
increasing amount of energy. As the tube clearly emerges above the photosphere and a strong 
current sheet forms, more action is seen by a fraction of the particles.
In \Fig{Trace.fig} two snapshots are shown. Here the box represents the 
full MHD domain. 
The field lines indicate the structure of the magnetic field, while the colour of the
particles indicates the energy obtained after 0.5 seconds of tracing.
Looking at the images it is seen that only a small part of the volume becomes energized and 
that these particles are accelerated away from the reconnection region following the global 
structure of the magnetic field. What is not clear from these images is the shape of the 
particle energy distribution function as they are channeled into the photosphere, where they 
are assumed to be de-accelerated almost instantaneously, creating strong bremsstrahlung 
radiation (we will return to this below). 

\begin{figure}
 {\hfill  \includegraphics[width=0.45\textwidth]{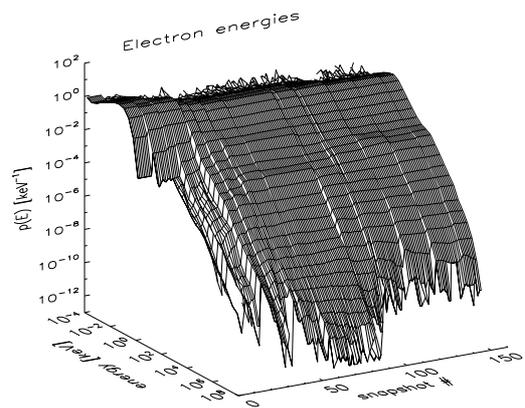} \hfill}

  \caption[]{\label{dist_function.fig} The figure shows the time evolution of the particle 
  kinetic energy distribution as a function of time in the numerical MHD experiment. 
  }
\end{figure}

It is clear from the images in \Fig{Trace.fig} that only a small fraction of the particles 
are exposed to a very strong acceleration (larger than 1 MeV). To make this clear, one can 
derive an energy distribution function after the 0.5 second integration. Doing this we 
ignore the fact that a large fraction of the particles that reach the defined boundaries do it
well before the the 0.5 second has been reached. These are included as having the
energy at the time of exit of the domain. \Fig{dist_function.fig} shows the time evolution 
of the kinetic energy distribution function. Initially we see the Maxwell distribution. As 
time in the numerical MHD experiment progresses a clear high-energy tail develops.  Only minor 
adjustments occur between the different MHD snapshots. The power slope of the distribution 
function is roughly maintained, while the peak energy changes, dependent on the
level of activity in the snapshot. Looking at a single distribution function one can
see different power slopes along the distribution, and it is possible to identify
three different power indexes.

The peak energy reached in the experiment is orders of magnitude larger than observed in 
solar flares. This same observation is noticed in a previous analysis 
\citep{2006A&A...449..749T}.
We will return to this in the discussion below.

\subsubsection{Impact structures}

\begin{figure}
  {\hfill \includegraphics[width=0.38\textwidth]{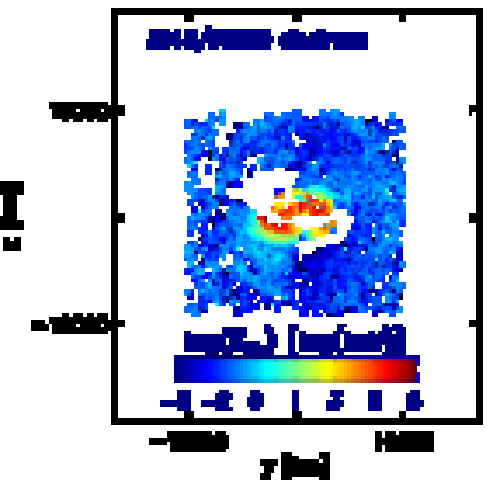} \hfill}

  {\hfill \includegraphics[width=0.38\textwidth]{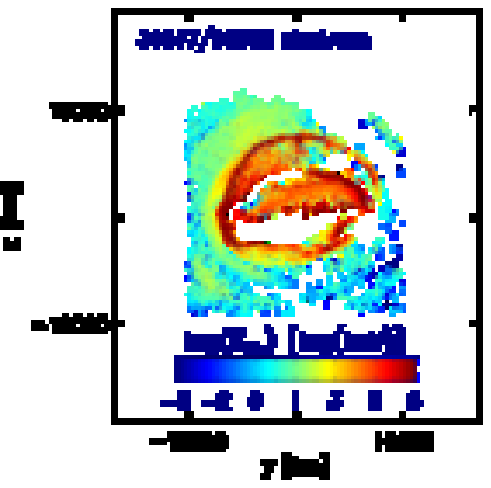} \hfill}

  {\hfill \includegraphics[width=0.38\textwidth]{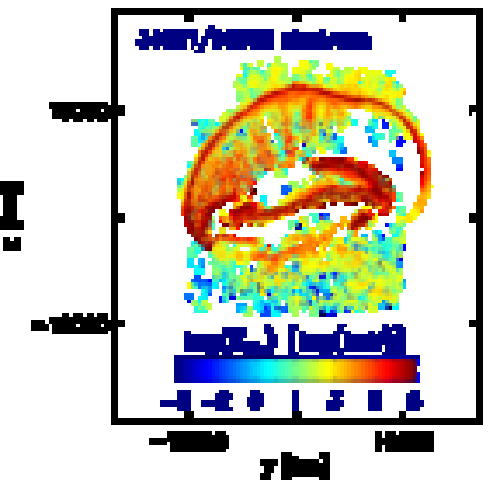} \hfill}

  \caption[]{\label{foot_point.fig} The pattern and energy distribution of the particles 
  after 0.5 seconds time integration. The colour indicates the particle kinetic energy
  on impact with the photosphere. The three images represent different times in the numerical MHD 
  experiment.}
\end{figure}

Direct observations of particle paths in the solar atmosphere is not possible. What is 
observed is the effect as the particles impact with the dense plasma in the 
photosphere. The radiation indicates where the field lines on which the accelerated particles 
move intersect with the photosphere. From the tracing experiments it is simple to make 
images showing where and
with which energy the particles exit the allowed tracing domain. The most direct comparison
can be made in the photosphere, were fast particles will loose their kinetic energy in 
exchange for heating the local plasma in a fraction of the observation time. 
This gives rise to local hot 
spots seen as $H_{\alpha}$ ribbons in flares. We plot the positions
and energies of the particles that reach the photosphere in \Fig{foot_point.fig}. The
frames show three snapshots representing different times in the numerical MHD experiment. The 
colour represents the energies of the particles impacting with the photosphere
at any time in the 0.5 second integration time. The numbers in the lower left of the 
frames indicate the number of particles (out of the total number) that impact with the 
photosphere.
The frames show a number of interesting features. In the early phases only few and low 
energy particles impact. As the reconnection sets in a characteristic pattern develops,
an almost closed envelope with a thick finger running down through part of the region. What 
does this signature express? And why do the later structures show much more complexity 
at especially one of the outer edges? These questions need to be answered in order to
better understand the acceleration process.

From the topology analysis of the numerical MHD experiment \citep[See][for details]{Galsgaard_09} it is
found that the reconnection process creates three independent flux domains. One represents 
the initially open magnetic field region, one confines the emerging magnetic flux and a 
final region represents the new loop system created by the ongoing magnetic 
reconnection process.  \Fig{connectivity.fig} above shows images of how the connectivity 
pattern of the numerical MHD experiment changes with time in the photosphere, as already discussed.

Comparing \Fig{connectivity.fig} and \Fig{foot_point.fig} reveals close similarities between 
the patterns of the two analyses. This is expected, as the region where particles are 
accelerated coincides with the location of the intersection of the separatrix surfaces. On 
one hand the separatrix is a mathematical surface defining the boundary between two flux 
regimes. The intersection of the separatrix surface with the photosphere therefore defines 
a mathematical line. On the other hand, the current sheet, and the associated diffusion 
region, has a finite spatial extension. The field lines 
penetrating the diffusion region therefore cover a much larger area when traced down to 
the photosphere. The spatial extent of the accelerated particles, when impacting with the
photosphere, must have a finite extent centered on the separatrix surfaces intersection with 
the photosphere. 

The structure and location of the particle impact with the photosphere is therefore a direct 
consequence of the extent and mapping for field lines penetrating the diffusion region 
and connecting to the photosphere. But, why does it include these small structural variations 
seen in the last of the images?  Looking at the same type of impact map further up in the 
atmosphere, the mapping contains a much simpler mapping structure. The reason for the 
complicated patterns is located in the lower parts of the atmosphere.  Here the magnetic 
fields lines are found to bend in different directions due to a weak convection like motion 
taking place. This indicates that the pressure gradient locally dominates the momentum 
equation, implying that the plasma density may be high enough to thermalize the accelerated 
particles at this height in the atmosphere. 


From comparing the images in the two figures it is clear that the impact region maps directly 
back to the acceleration region. Being able to make realistic field extrapolations from 
magnetograms into the corona, will therefore give a clear indication of where the diffusion 
region will be located in 3D space. Doing this with large precision is not a simple task in
a magnetic field that is far from being in a potential or even linear force free state.

\subsection{Radiation spectra from the photospheric impact}

From a physical point of view the distribution functions give direct insight to the 
behavior of the particles. But, it is not a quantity that can be easily observed. Instead, 
we observe spectra from the radiation particles emit as they are de-accelerated on
impact with the photosphere. This process occurs almost instantaneously
which allows us to calculate the spectra directly from the particle distribution function. 
\citet{1971SoPh...18..489B} and \cite{2003ApJ...586..606H} provide detailed information 
about how to do this. Because of
the very high energy tail of the distribution function, the relativistic formulae for the
derivation must be used. This limits the energy of the most energetic photons in the
spectrum, as it would otherwise only show one power law extending to very high energies. 
\begin{figure}
  {\hfill \includegraphics[width=0.45\textwidth]{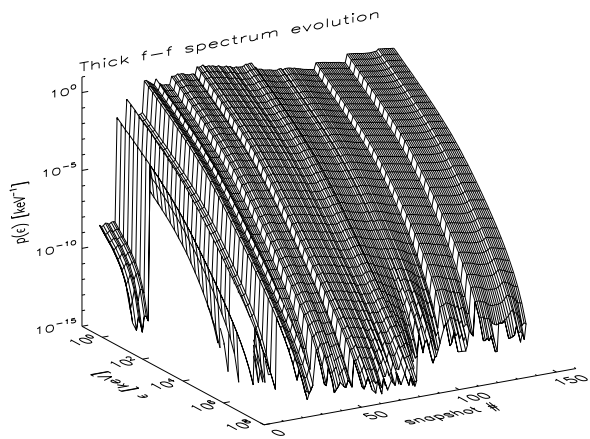} \hfill} 
  \caption[]{\label{spectra.fig} The spectra obtained directly from the electron kinetic 
   energy distribution functions for the electrons impacting with the photosphere in the 
   tracing experiments. The assumption is that the particles are thermalised over a
   short timescale (compared with the observation time) and distance through bremsstrahlung.}
\end{figure}

The spectra obtained from this exercise is shown in \Fig{spectra.fig}. This shows two 
situations, both containing three different power law spectra depending on the energy range of 
the spectrum. The slope of the spectrum, covering the X-ray to gamma-ray range, changes from
about 1.15 and 1.90. These slopes represent much harder spectra than observed from
flare events. For large flares, the peak energy typically reaches hundreds of MeV and the 
X-ray power index is typically in the range of 2.6 to 3.9 \citep{2008AIPC.1039...52L}.

\section{Discussion}  
\label{Discussion.sec}
The previous section discusses the results we have obtained tracing test particles using
MHD snapshots from a numerical flux emergence experiment. We have found that even in a situation where 
there is only one large current sheet located inside the numerical domain, is it possible to 
accelerate particles away from the initially given Maxwell distribution. The final result is 
an electron distribution function that contains a high energy power law tail which extends 
from the Maxwell 
distribution. After the onset of MHD reconnection, this distribution forms rapidly and 
maintains a similar structure throughout, although the peak energy and fraction of accelerated
particles varies with time. There are a number of critical issues related to the approach 
used here: 

\begin{itemize}
\item{
Individual particles follow paths that gyrate around magnetic field lines. As part of the 
analysis we have compared the results obtained using both the correct Lorentz equation for 
the particle motion and the much faster GCA. It is clear that for this experiment, the 
two methods produce almost identical results. The reason for this comes from the fact
that the background field from the numerical MHD experiment does not have a grid resolution that is 
even close to the Larmor radius of the particle orbits. The magnetic and electric field 
therefore have spatial variations on length scales that are many orders of magnitude larger, 
which makes the GCA a good description of the particle motion. There is therefore no reason 
to use the LE description as long as the numerical MHD experiment describes phenomena on typical solar 
surface length scales.
}
\item{
A limitation using the test particle tracing is the lack of interaction between the
particles present in the domain. Each particle is treated independently of all other
particles in the domain. This has a number of problems:
  \begin{itemize}
  \item{
  There are no built in collisions, which allows particles to be accelerated far more than 
  in a realistic plasma.
  }
  \item{
  There is no electric interaction, implying that the charge separation between the 
  electrons and the much slower protons is again unrealistic.
  }
  \end{itemize} 
}
\item{
Even if we included simple particle collisions, then a number of particles would reach high
energies and their collision cross sections would decrease sufficiently for them to escape 
the acceleration region. But, as the particles are accelerated, there is a natural exchange 
of energy with the macroscopic field. This would decrease the E-field in the diffusion 
region and the remaining particles would feel a smaller acceleration force, making it less 
likely that they would experience the same significant acceleration and escape from the 
region. 
}
\item{
The above items have significant implications on the obtained distribution functions. In the
present approach too many particles reach far too high energies. The lack
of back reaction on the macroscopic fields allows the production of power laws that are 
too hard compared with observations. These problems naturally influence the derived 
bremsstrahlung spectrum generated by the particles impacting with the photosphere.
}
\end{itemize}

To improve on these issues we need a different and much more calculation heavy approach. 
One way is to adopt a PIC approach, where one assumes the included particles to have a 
very exaggerated weight to allow the domain to cover the MHD scales without being 
dependent on resolving spatially the Larmor radii of real charged particles. This 
approach is being investigated.  

\section{Conclusions}
\label{Conclusions.sec}

The usage of test particle tracking in MHD simulations constitutes a simple way to 
obtain information about regions of particle acceleration and the spectra and spatial 
distribution on planes of interest.  When the MHD data represents a large solar structure 
and the typical length scale involved is much larger than the particle Larmor radius, 
the Guiding Center Approximation is the favored method to obtain these results. 
Comparisons between the GCA and the full Lorentz solution for this data discussed in 
this paper show no differences in results, while on the computational side the GCA is 
between a 100 and a 1000 time cheaper in computing time.  

Warnings must be given. The lack of back reactions between the accelerated particles and the 
provided MHD field results in far too many particles being accelerated to far too high 
energies. The distribution functions and spectra obtained from such calculations are 
therefore giving too many high energy particles resulting in too hard spectra when calculating 
the bremsstrahlung emitted by the particles interacting with the dense photosphere.

On the positive side, it is possible to use the impact location on the photosphere to 
identify the location of the separator surfaces connecting up to the diffusion region where 
the acceleration is taking place. This information may be a help to guide magnetic field 
extrapolations of the photospheric
field distribution, as one can both see the impact points in the photosphere and the X-ray 
region in the corona where the reconnection process takes place. These have to fit with the 
topological structure obtained from the extrapolation models.

To obtain more generally useful information from this type of exercise one has to adopt a 
more physically consistent model, which takes the whole approach into a much more complicated
problem. But one we have to find a solvable solution to
if we want to obtain reliable information from numerical experiments.

\bibliographystyle{aa} 
\bibliography{13541ref.bib} 

\begin{thebibliography}{29}
\expandafter\ifx\csname natexlab\endcsname\relax\def\natexlab#1{#1}\fi

\bibitem[{{Ambrosiano} {et~al.}(1988){Ambrosiano}, {Matthaeus}, {Goldstein}, \&
  {Plante}}]{1988JGR....9314383A}
{Ambrosiano}, J., {Matthaeus}, W.~H., {Goldstein}, M.~L., \& {Plante}, D. 1988,
  \jgr, 93, 14383

\bibitem[{{Birn} {et~al.}(2005){Birn}, {Galsgaard}, {Hesse}, {Hoshino}, {Huba},
  {Lapenta}, {Pritchett}, {Schindler}, {Yin}, {B{\"u}chner}, {Neukirch}, \&
  {Priest}}]{2005GeoRL..3206105B}
{Birn}, J., {Galsgaard}, K., {Hesse}, M., {et~al.} 2005, \grl, 32, 6105

\bibitem[{{Birn} \& {Priest}(2007)}]{2007rmfm.book.....B}
{Birn}, J. \& {Priest}, E.~R. 2007, {Reconnection of magnetic fields :
  magnetohydrodynamics and collisionless theory and observations}, ed.
  J.~{Birn} \& E.~R. {Priest}

\bibitem[{{Brown}(1971)}]{1971SoPh...18..489B}
{Brown}, J.~C. 1971, \solphys, 18, 489

\bibitem[{{Dmitruk} \& {Matthaeus}(2006)}]{2006JGRA..11112110D}
{Dmitruk}, P. \& {Matthaeus}, W.~H. 2006, Journal of Geophysical Research
  (Space Physics), 111, 12110

\bibitem[{{Dmitruk} {et~al.}(2004){Dmitruk}, {Matthaeus}, \&
  {Seenu}}]{2004ApJ...617..667D}
{Dmitruk}, P., {Matthaeus}, W.~H., \& {Seenu}, N. 2004, \apj, 617, 667

\bibitem[{{Dmitruk} {et~al.}(2003){Dmitruk}, {Matthaeus}, {Seenu}, \&
  {Brown}}]{2003ApJ...597L..81D}
{Dmitruk}, P., {Matthaeus}, W.~H., {Seenu}, N., \& {Brown}, M.~R. 2003, \apjl,
  597, L81

\bibitem[{{Galsgaard} \& {Moreno-Insertis}(2009)}]{Galsgaard_09}
{Galsgaard}, K. \& {Moreno-Insertis}, F. 2009, In preperation

\bibitem[{{Holman}(2003)}]{2003ApJ...586..606H}
{Holman}, G.~D. 2003, \apj, 586, 606

\bibitem[{{Hoyng} {et~al.}(1981){Hoyng}, {Duijveman}, {Machado}, {Rust},
  {Svestka}, {Boelee}, {de Jager}, {Frost}, {Lafleur}, {Simnett}, {van Beek},
  \& {Woodgate}}]{1981ApJ...246L.155H}
{Hoyng}, P., {Duijveman}, A., {Machado}, M.~E., {et~al.} 1981, \apjl, 246,
  L155+

\bibitem[{{Isobe} {et~al.}(2002){Isobe}, {Yokoyama}, {Shimojo}, {Morimoto},
  {Kozu}, {Eto}, {Narukage}, \& {Shibata}}]{2002ApJ...566..528I}
{Isobe}, H., {Yokoyama}, T., {Shimojo}, M., {et~al.} 2002, \apj, 566, 528

\bibitem[{{Jing} {et~al.}(2007){Jing}, {Lee}, {Liu}, {Gary}, \&
  {Wang}}]{2007ApJ...664L.127J}
{Jing}, J., {Lee}, J., {Liu}, C., {Gary}, D.~E., \& {Wang}, H. 2007, \apjl,
  664, L127

\bibitem[{{Krucker} {et~al.}(2008){Krucker}, {Hurford}, {MacKinnon}, {Shih}, \&
  {Lin}}]{2008ApJ...678L..63K}
{Krucker}, S., {Hurford}, G.~J., {MacKinnon}, A.~L., {Shih}, A.~Y., \& {Lin},
  R.~P. 2008, \apjl, 678, L63

\bibitem[{{Lin}(2008)}]{2008AIPC.1039...52L}
{Lin}, R.~P. 2008, in American Institute of Physics Conference Series, Vol.
  1039, American Institute of Physics Conference Series, ed. G.~{Li}, Q.~{Hu},
  O.~{Verkhoglyadova}, G.~P. {Zank}, R.~P. {Lin}, \& J.~{Luhmann}, 52--62

\bibitem[{{Lin} \& {Hudson}(1976)}]{1976SoPh...50..153L}
{Lin}, R.~P. \& {Hudson}, H.~S. 1976, \solphys, 50, 153

\bibitem[{{Lin} {et~al.}(2003){Lin}, {Krucker}, {Hurford}, {Smith}, {Hudson},
  {Holman}, {Schwartz}, {Dennis}, {Share}, {Murphy}, {Emslie}, {Johns-Krull},
  \& {Vilmer}}]{2003ApJ...595L..69L}
{Lin}, R.~P., {Krucker}, S., {Hurford}, G.~J., {et~al.} 2003, \apjl, 595, L69

\bibitem[{{Lin} \& {Rhessi Team}(2002)}]{2002ESASP.506.1035L}
{Lin}, R.~P. \& {Rhessi Team}. 2002, in ESA Special Publication, Vol. 506,
  Solar Variability: From Core to Outer Frontiers, ed. A.~{Wilson}, 1035--1044

\bibitem[{{Liu} {et~al.}(2008{\natexlab{a}}){Liu}, {Lee}, {Jing}, {Gary}, \&
  {Wang}}]{2008ApJ...672L..69L}
{Liu}, C., {Lee}, J., {Jing}, J., {Gary}, D.~E., \& {Wang}, H.
  2008{\natexlab{a}}, \apjl, 672, L69

\bibitem[{{Liu} \& {Wang}(2009)}]{2009ApJ...696L..27L}
{Liu}, C. \& {Wang}, H. 2009, \apjl, 696, L27

\bibitem[{{Liu} {et~al.}(2008{\natexlab{b}}){Liu}, {Petrosian}, {Dennis}, \&
  {Jiang}}]{2008ApJ...676..704L}
{Liu}, W., {Petrosian}, V., {Dennis}, B.~R., \& {Jiang}, Y.~W.
  2008{\natexlab{b}}, \apj, 676, 704

\bibitem[{{Masuda} {et~al.}(1994){Masuda}, {Kosugi}, {Hara}, {Tsuneta}, \&
  {Ogawara}}]{1994Natur.371..495M}
{Masuda}, S., {Kosugi}, T., {Hara}, H., {Tsuneta}, S., \& {Ogawara}, Y. 1994,
  \nat, 371, 495

\bibitem[{{Moreno-Insertis} {et~al.}(2008){Moreno-Insertis}, {Galsgaard}, \&
  {Ugarte-Urra}}]{2008ApJ...673L.211M}
{Moreno-Insertis}, F., {Galsgaard}, K., \& {Ugarte-Urra}, I. 2008, \apjl, 673,
  L211

\bibitem[{Nordlund \& Galsgaard(1995)}]{Nordlund+Galsgaard95mhd}
Nordlund, {\AA}. \& Galsgaard, K. 1995, A 3D {MHD} Code for Parallel Computers,
  Tech. rep., Astronomical Observatory, Copenhagen University

\bibitem[{{Northrop}(1964)}]{1964AmJPh..32..807N}
{Northrop}, T.~G. 1964, American Journal of Physics, 32, 807

\bibitem[{{Petschek}(1964)}]{1964NASSP..50..425P}
{Petschek}, H.~E. 1964, NASA Special Publication, 50, 425

\bibitem[{{Press} {et~al.}(1992){Press}, {Teukolsky}, {Vetterling}, \&
  {Flannery}}]{1992nrfa.book.....P}
{Press}, W.~H., {Teukolsky}, S.~A., {Vetterling}, W.~T., \& {Flannery}, B.~P.
  1992, {Numerical recipes in FORTRAN. The art of scientific computing}
  (Cambridge: University Press, |c1992, 2nd ed.)

\bibitem[{{Priest} \& {Forbes}(2000)}]{2000mare.book.....P}
{Priest}, E. \& {Forbes}, T. 2000, {Magnetic Reconnection}, ed. E.~{Priest} \&
  T.~{Forbes}

\bibitem[{{Shibata} {et~al.}(2007){Shibata}, {Nakamura}, {Matsumoto}, {Otsuji},
  {Okamoto}, {Nishizuka}, {Kawate}, {Watanabe}, {Nagata}, {UeNo}, {Kitai},
  {Nozawa}, {Tsuneta}, {Suematsu}, {Ichimoto}, {Shimizu}, {Katsukawa},
  {Tarbell}, {Berger}, {Lites}, {Shine}, \& {Title}}]{2007Sci...318.1591S}
{Shibata}, K., {Nakamura}, T., {Matsumoto}, T., {et~al.} 2007, Science, 318,
  1591

\bibitem[{{Turkmani} {et~al.}(2006){Turkmani}, {Cargill}, {Galsgaard},
  {Vlahos}, \& {Isliker}}]{2006A&A...449..749T}
{Turkmani}, R., {Cargill}, P.~J., {Galsgaard}, K., {Vlahos}, L., \& {Isliker},
  H. 2006, \aap, 449, 749

\end{thebibliography}
\end{document}